\documentclass[aps,twocolumn]{revtex4} 
\usepackage{psfig}
\usepackage{bm}

\begin{document}

\title{Fluctuation-Dissipation Ratio for Compacting Granular Media} 
 
\author{Alain Barrat$^1$, Vittoria Colizza$^{2,3}$ and Vittorio
Loreto$^3$}

\affiliation{$^1$ Laboratoire de Physique Th{\'e}orique, Unit{\'e}
Mixte de Recherche UMR 8627, B{\^a}timent 210, Universit{\'e} de
Paris-Sud, 91405 Orsay Cedex, France\\$^2$ International School for
Advanced Studies (SISSA), via Beirut 4, 34014 Trieste, Italy\\$^3$
``La Sapienza'' University in Rome, Physics Department, P.le A. Moro
5, 00185 Rome, Italy and INFM, Center for Statistical Mechanics and
Complexity, Rome, Italy}

\date{\today}

\begin{abstract} 
In this paper we investigate the possibility of a dynamical definition
of an effective temperature for compacting granular media in the
framework of the Fluctuation-Dissipation (FD) relations. We have
studied two paradigmatic models for the compaction of granular media,
which consider particles diffusing on a lattice, with either
geometrical (Tetris model) or dynamical (Kob-Andersen model)
constraints. Idealized compaction without gravity has been implemented
for the Tetris model, and compaction with a preferential direction
imposed by gravity has been studied for both models.  In the ideal
case of an homogeneous compaction, the obtained FD ratio is clearly
shown to be in agreement with the prediction of Edwards' measure at
various densities. Similar results are obtained with gravity only when
the homogeneity of the bulk is imposed. In this case the FD ratio
obtained dynamically for horizontal displacements and mobility and
from Edwards' measure coincide.  Finally, we propose experimental
tests for the validity of the Edwards' construction through the
comparison of various types of dynamical measurements.  (PACS:
05.70.Ln, 05.20.-y, 45.70.Cc)
\end{abstract} 

\maketitle

\section{Introduction}

Granular materials~\cite{Nagel,Nagel2,grain} play a very important
role in many fields of human life and industrial activities, like
agriculture, building, chemistry, etc. Their properties are
interesting not only for practical reasons, but also from the point of
view of fundamental physics. In fact, in spite of their apparent
simplicity, they display a wide variety of behaviours that is only
partially understood in terms of general physical principles.

\subsection{Typical problems to face in the study of compact granular matter}

The common wisdom about granular materials defines them as non-thermal
systems, since thermal energy can be generally neglected if compared
to mechanical energy due to gravity and other external energy sources
usually acting on these systems. In addition, a fundamental role in
the dynamics is played by the mechanical energy dissipation due to
friction and collisions among the grains and with the container walls:
motion can take place only by continuously feeding energy into the
system, that otherwise would get stuck into some metastable state, no
longer exploring spontaneously the configuration space. Consequently,
as a matter of fact, the dynamics of granular matter is always a
response to an external perturbation and in general the response will
depend in a non-trivial way on the rheological properties of the
medium, on the boundaries, on the driving procedure and, last but not
the least, on the past history of the system.

From the non-thermal character of these systems a lot of consequences
can be drawn. The first one concerns the lack of any ergodicity
principles: a granular media is not able to freely explore its phase
space and the dynamical equations do not leave the microcanonical or
any other known ensemble invariant. Moreover, just as in the case of
aging glasses, the compaction
dynamics~\cite{exp-chicago,exp-rennes,exp-nicolas} does not approach
any stationary state on experimental time scales (at least at small
enough forcing) and these systems exhibit
aging~\cite{nicodemi_aging,BaLo,talbot} and
memory~\cite{Josserand,BaLo2,brey_memory}. A granular media lives then
always in non-equilibrium conditions. Even when one observes some
stationary state as the result of a specific dynamics (energy
injection) imposed to the system, one is never able to establish some
sort of equipartition principle ruling how the energy injected is
redistributed among the different degrees of freedom of the system.

Despite all these difficulties, since granular systems involve a large
number of particles, one is always naturally inclined to treat them
with methods of statistical mechanics. In particular one of the main
questions concerns the very possibility to construct a coherent
thermodynamics and on this point the debate is wide open. The
construction of a thermodynamics would imply the identification of a
suitable distribution that is left invariant by the dynamics ({\em
e.g.} the microcanonical ensemble), and then the assumption that this
distribution will be reached by the system, under suitable conditions
of 'ergodicity'. As already mentioned since in granular media energy
is lost through internal friction, and gained by a non-thermal source
such as tapping or shearing, this approach seems doomed from the
outset.  Nevertheless one could ask whether some elementary
thermodynamic quantities are well defined and what is their meaning.
It is in this spirit that in this paper we address the question of the
definition of an effective temperature for compact granular media.
Before going in the details let us briefly review the state of the art
on these subjects.

\subsection{State of the art}

A lot of approaches~\cite{herrmann}-\cite{entropy} have been devised
in the last years to provide a coherent scenario but till now the
situation is quite uncertain mainly because at a fundamental level
there is no general argument showing that a particular construction
should lead to the relevant distribution for the dynamics (as one does
in the case of conservative dynamics, leading to thermodynamics).

Many models have been proposed in order to reproduce the rich
phenomenology observed in granular compaction
experiments~\cite{exp-chicago,exp-rennes}, but a general
thermodynamical-like framework, based on the idea of describing
granular material with a small number of parameters, is however still
lacking.

A very ambitious approach, aiming at such a description of dense
granular matter has been put forward by S. Edwards and
co-workers~\cite{edwards,anita}, by proposing an equivalent of the
microcanonical ensemble: macroscopic quantities in a jammed situation
should be obtained by a flat average over all {\em blocked
configurations} (i.e. in which every grain is unable to move) of given
volume, energy, etc.. The strong assumption is here that all blocked
configurations are treated as equivalent and have the same weight in
the measure.

Very recently, important progresses in this direction have been
reported in various contexts: a tool to systematically construct
Edwards' measure, defined as the set of blocked configurations of a
given model, was proposed in~\cite{bakulose1,bakulose2}; it was used
to show that the outcome of the aging dynamics of the Kob-Andersen
(KA)~\cite{KoAn} model (a kinetically constrained lattice gas model)
was correctly predicted by Edwards' measure. Moreover, the validity
and relevance of Edwards' measure have been demonstrated~\cite{CoLoBa}
for the Tetris model~\cite{prltetris}, for one-dimensional
phenomenological models~\cite{Brey}, for spin models with ``tapping''
dynamics~\cite{Dean}, and for sheared hard spheres~\cite{Makse}.

At the time being however, the correspondence between Edwards'
distribution and long-time dynamics is at best checked but does not
follow from any principle.  It is therefore important to continue to
explore its range of validity and therefore to test its applicability
to various kinds of models.

Another important message emerging from these studies concerns the
link between Edwards' approach and the outcomes of the measurements of
the Fluctuation-Dissipation
relations. In~\cite{bakulose1,bakulose2,CoLoBa} it has been shown in
the framework of two non-mean field models, the Kob-Andersen model and
the Tetris model, that the so-called Edwards' ratio (see below for its
precise definition) coincides on a wide range of densities with the
Fluctuation-Dissipation Ratio (FDR) in homogeneous systems, i.e. in
systems without any preferential direction. This paper extends those
results providing a series of new evidences for the validity of
Edwards' approach in two main directions. {\bf (i)} First of all we
focus on more realistic situations by considering the case of granular
packings subject to gravity: this is an important example to test the
role of large scale inhomogeneities, such as the density profiles
along the preferential direction, whose treatment has to be performed
very carefully in order to avoid apparently non-physical
results~\cite{nicodemi_response}. {\bf (ii)} We present new results
concerning the independence of the FD ratio of the observables used
for its definition.

It should be noted that previous measurements of FDR 
in the presence of gravity (and thus heterogeneities) have
been attempted in~\cite{nicodemi_response,Se2}. 
However, the negative response functions observed in~\cite{nicodemi_response}
was subsequently shown in~\cite{BaLo,BaLo2} to be linked to memory 
effects~\cite{Josserand}, and not to Edwards' measure. Moreover, 
the measures of~\cite{nicodemi_response,Se2} were flawed 
by an incorrect definition of the correlation part
of the Fluctuation-Dissipation ratio, due to the fact that 
one-time quantities are still evolving
(see section~\ref{sec:V} for a detailed discussion). The
conclusion of~\cite{Se2} about the
existence of a dynamical temperature was thus premature.

The link established between Edwards' approach and the
Fluctuation-Dissipation relations could open new perspectives from two
different points of view.  First of all from the experimental point of
view where possibilities are open to (i) check the Edwards' measure by
means of dynamical measurements; (ii) perform dynamical measurements
(through the Fluctuation-Dissipation relations) of a ``temperature''
which should only depend on the density. This could allow for an at
least partial equilibration of the disproportion existing between the
huge number of theoretical/numerical works (and this paper contributes
to this number) and the few experimental results. Moreover very
focused experimental results could help in discriminating between the
different models proposed in literature.  On the other hand, from the
theoretical point of view one is left with several questions: why
Edwards' measure seems to be correct in a wide range of situations?
Is it possible to identify some first principles justifications or
derivations for it?  Why the outcomes of the Edwards' approach seem to
coincide with the results of Fluctuation-Dissipation measurements?

This paper, far from being able to address all these questions, tries
to make the link between Edwards' approach and the
Fluctuation-Dissipation measurements firmer in several realistic
situations and propose some possible experimental paths for its
checking. The outline of the paper is as follows: we first recall in
section~\ref{sec:II} the definition of the models under consideration,
and in section~\ref{sec:III} how to construct Edwards' measure. The
case of homogeneous compaction for the Tetris model is described in
section~\ref{sec:IV}, while fluctuation-dissipation ratios during a
gravity-driven compaction are measured for both KA and Tetris models
in section~\ref{sec:V}.  Finally, possible experiments are proposed in
section~\ref{sec:VI}, and conclusions are drawn.

\section{Models definition}
\label{sec:II}

The models we consider are lattice models, and in this sense are not
realistic microscopic models of granular materials. However, they are
worth investigating: on the one hand, they have been shown to
reproduce the complex phenomenology of granular media (see
\cite{nicodemi_aging,BaLo,BaLo2,prltetris} and \cite{KuPeSe,SeAr}); on
the other hand, the validity of Edwards' measure for some observables
has already been shown in an ideal case of homogeneous compaction
\cite{bakulose1,bakulose2}, making these finite-dimensional models
good candidates for further investigations in more realistic
situations, i.e. with heterogeneities induced by gravity and the
existence of a preferential direction.

\subsection{Tetris model}

Under the denomination of ``tetris'' falls a class of
lattice~\cite{prltetris} models whose basic ingredient is the
geometrical frustration. The models are defined on a two-dimensional
square lattice with particles of randomly chosen shapes and sizes. The
only constraint in the system is that particles cannot overlap: for
two nearest-neighbor particles, the sum of the arms oriented along the
bond connecting the two particles has to be smaller than the bond
length. The interactions are hence not spatially quenched but
determined in a self-consistent way by local particle configurations.

In the version we use (see~\cite{bakulose2}), the particles have a
``T''-shape (three arms of length $\frac{3}{4} d$, where $d$ sets the
bond size on the square lattice).  The maximum density allowed is then
$\rho_{max}=2/3$.

\subsection{Kob-Andersen model}

The other model we consider is the so-called Kob-Andersen (KA)
model~\cite{KoAn}, first studied in the context of Mode-Coupling
theories~\cite{gotze} as a finite dimensional model exhibiting a
divergence of the relaxation time at a finite value of the control
parameter (here the density); this divergence is due to the presence
in this model of the formation of ``cages'' around particles at high
density (the model was indeed devised to reproduce the cage effect
existing in super-cooled liquids).

Though very schematic, it has then been shown to reproduce rather well
several aspects of glasses~\cite{KuPeSe}, like the aging behaviour
with violation of FDT~\cite{Se}, and of granular
compaction~\cite{SeAr}.

The successful comparison of aging dynamics and predictions of
Edwards' measure was moreover shown for the first time for this model,
in~\cite{bakulose1,bakulose2}, in the idealized case of homogeneous
compaction.  On the other hand, a study of the violation of the FDT
during the compaction process (under gravity) was undertaken
in~\cite{Se2}, and the existence of a dynamical temperature was
advocated~\footnote{we will see in section~\ref{sec:V} that the
situation is however more complex and that great care has to be taken
when measuring FDT violations under gravity.}.

The model is defined as a lattice gas on a three dimensional lattice,
with at most one particle per site. The dynamical rule is as follows:
a particle can move to a neighboring empty site, only if it has
strictly less than $\nu$ neighbours in the initial and in the final
position.

Following~\cite{KoAn}, we take $\nu=5$: this ensures that the system
is still ergodic at low densities, while displaying a sharp increase
in relaxation times at a density well below $1$.  The dynamic rule
guarantees that the equilibrium distribution is trivially simple since
all the configurations of a given density are equally probable: the
Hamiltonian is just $0$ since no static interactions exist.

Moreover, it is also easy to consider a mixture of
two types of particles, by considering particles
of type $1$ with a certain value $\nu_1$ for the dynamical constraint,
and particles of type $2$ with $\nu_2 \ne \nu_1$~\cite{SeAr}.

\section{Edwards' measure}
\label{sec:III}

Edwards' approach is based on a flat sampling over all the blocked
configurations, i.e. configurations with all particles unable to
move. This definition therefore depends on the model and of the type
of dynamics. For example, a particle is more easily blocked in
presence of an imposed drift, e.g. gravity.

This approach, based on the idea of describing granular material with
{\em a small number of parameters}, leads to the introduction of an
entropy $S_{Edw}$, given by the logarithm of the number of blocked
configurations of given volume, energy, etc., and its corresponding
density $s_{Edw}\equiv S_{Edw}/N$.  Associated with this entropy are
the state variables such as `compactivity'
$X_{Edw}^{-1}=\frac{\partial}{\partial V}S_{Edw}(V)$ and `temperature'
$T_{Edw}^{-1}=\frac{\partial}{\partial E}S_{Edw}(E)$.

The explicit construction of Edwards' measure, as well as of the
equilibrium measure, has been described in detail
in~\cite{bakulose1,bakulose2} for the Tetris and the KA models.  In
particular, Edwards' measure is obtained with an annealing procedure
at fixed density. In order to select only the subset of configurations
contributing to the Edwards' measure we introduce an auxiliary
temperature $T_{aux}$ (and the corresponding $\beta_{aux}=1/T_{aux}$)
and, associated to it, an auxiliary energy $E_{aux}$ which, for each
configuration, is equal to the number of mobile particles. A particle
is defined as mobile if it can be moved according to the dynamic rules
of the original model.

In particular one measures $E_{aux}(\beta_{aux},\rho)$, i.e. the
decrease of the auxiliary energy at fixed density, performing an
annealing procedure increasing progressively $\beta_{aux}$. From this
measure one can compute the Edwards' entropy density defined by:

\begin{eqnarray}
s_{Edw}(\rho) & \equiv & s_{aux}(\beta_{aux}=\infty,\rho)= \nonumber
\\ && s_{equil}(\rho) - \int_0^\infty e_{aux}(\beta_{aux},\rho)
d\beta_{aux} \,
\label{form_entro_edw}
\end{eqnarray}
where $e_{aux}(\beta_{aux},\rho)$ is the auxiliary Edwards' energy
density and $s_{aux}(\beta_{aux}=0,\rho) = s_{equil}(\rho)$ is the
equilibrium entropy of the model.

For the KA model, the equilibrium entropy is simply the entropy of a
lattice gas.  It is worth however recalling that, for the Tetris model
(and in general when the equilibrium measure is not known
analytically), the equilibrium measure can also be obtained with an
annealing procedure: also in this case one introduces an auxiliary
temperature $T'_{aux}=1/\beta'_{aux}$ associated with an auxiliary
energy $E'_{aux}$ defined as the total particle overlaps existing in a
certain configuration. For each value of $T'_{aux}$ one allows the
configurations with a probability given by $e^{(-\beta'_{aux}
E'_{aux})}$.  Starting with a large temperature $T'_{aux}$ one samples
the allowed configurations by progressively decreasing $T'_{aux}$.  As
$T'_{aux}$ is reduced $E'_{aux}$ decreases and only at $T'_{aux}=0$
(no violation of constraints allowed) the auxiliary energy is
precisely zero. The exploration of the configuration space can be
performed working at constant density by interchanging the positions
of couples of particles. This procedure is used to compute
$E'_{aux}(\beta'_{aux},\rho)$ and $e'_{aux}(\beta'_{aux},\rho)$
(energy per particle), from which one can compute the equilibrium
entropy per particle by the expression:

\begin{eqnarray}
s_{equil}(\rho) & \equiv &
s_{equil}(\beta'_{aux}=\infty,\rho)=\nonumber\\ &= &
s_{equil}(\beta'_{aux}=0,\rho) - \int_0^\infty
e'_{aux}(\beta'_{aux},\rho) d\beta'_{aux} \,
\label{form_entro_equi}
\nonumber
\end{eqnarray}
where $e'_{aux}(\beta'_{aux},\rho)$ is the auxiliary energy per
particle.  For the choice made for the particles one has
\begin{eqnarray}
s_{equil}(\beta'_{aux}=0,\rho) & = &
-\rho\ln\rho-(1-\rho)\ln(1-\rho)\nonumber \\
&& +\rho\ln 4,
\end{eqnarray}
which is easily obtained by counting the number of ways in which one
can arrange $\rho L^2$ particles of four different types on $L^2$
sites.

\begin{figure}[htb] 
\centerline{\psfig{file=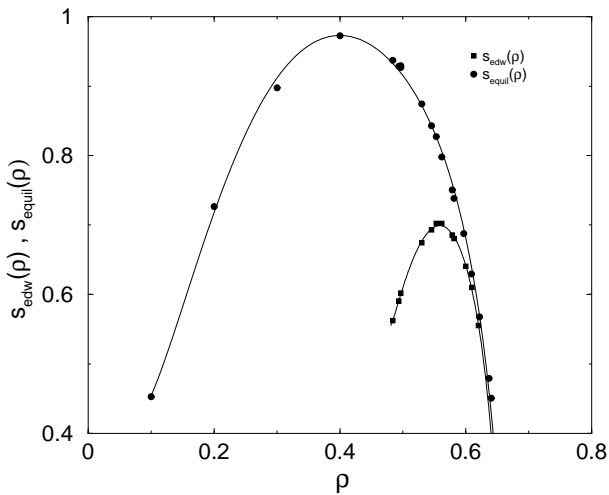,width=8.0cm,angle=-90}}
\vspace{0.2cm}
\caption{$s_{Edw}(\rho)$ and $s_{equil}(\rho)$ for the Tetris model.}
\label{fig:entro} 
\end{figure} 

Edwards' and equilibrium entropies, computed as a function of density,
are reported in Fig.~\ref{fig:entro} for the Tetris model.  It is also
possible to compute the so-called Edwards' ratio, defined as
\begin{equation}
X_{Edw}\,=\, \frac{ds_{Edw}(\rho)}{d \rho} \, \Big/ \,
\frac{ds_{equil}(\rho)}{d \rho} \ .
\end{equation}
$X_{Edw}$ approaches 1 as $\rho \to \rho_{max}$, since at the maximum
density all configurations become blocked and therefore equilibrium
and Edwards' entropies become equivalent.

\section{Fluctuation-Dissipation Ratio for the case without gravity}
\label{sec:IV}

\subsection{Tetris model: dynamics without gravity}

It is worth to recall that for the KA, the compaction dynamics was
obtained by means of a `piston', i.e. by creating and destroying
particles only on the topmost layer (of a cubic lattice of linear size
$L$) with a chemical potential $\mu$~\cite{KuPeSe}. The validity of
Edwards' measure in this case has been described
in~\cite{bakulose1,bakulose2}. We will therefore focus here on the
Tetris model. In~\cite{bakulose2,CoLoBa}, compaction dynamics without
gravity has been implemented for this model in order to avoid
generating a preferential direction.

The system is initialized through a random sequential adsorption of
``T''-shaped particles on an initially empty lattice. The grains must
satisfy the geometrical constraints with their nearest neighbours and
cannot diffuse on the lattice. This filling procedure stops when no
other particles can be deposited on the system anymore, yielding a
reproducible initial density of $\rho \approx 0.547$.

The irreversible compaction dynamics is then realized alternating
attempted random diffusions (in which a particle is chosen with
uniform probability and allowed to move toward one of its
nearest-neighbours with probability $p=\frac{1}{4}$, only if all the
possible geometrical constraints are satisfied), and attempted random
depositions on the lattice (an empty site is chosen with uniform
probability and a grain is then adsorbed on the lattice only if no
violation of the geometrical constraints occurs). The global density
increases, the system remaining homogeneous during the process.

The Monte Carlo time unit is defined as the number of elementary
dynamical steps normalized to the number of sites of the lattice,
$L^2$.  In order to overcome the problem related to the simulation of
very slow processes and obtain a reasonable number of different
realizations to produce clean data, we have devised a fast algorithm
(in the spirit of Bortz-Kalos-Lebowitz algorithm~\cite{bkl}), where
the essential ingredient is the updating of a list of mobile particles
(whose number is $n_{mob}$). In order to reduce the number of less
significant events, such as failed attempts of deposition/diffusion,
this algorithm is essentially based on a guided dynamics where only
mobile particles (i.e. grains which could diffuse toward a
neighbouring site) are considered. At each time step one {\em mobile}
particle is chosen with uniform probability and allowed to move if all
the geometrical constraints are satisfied. If the attempt has been
successful, the list of mobile particles is then updated, performing a
\emph{local} control of grains' mobility. This procedure therefore
introduces a temporal bias in the evolution of the system, which has
to be taken into account by incrementing the time of an amount $\Delta
t= 1/n_{mob}$, after each guided elementary step. This algorithm
becomes very efficient as the density of the system increases, since
the number of mobile particles reduces drastically.

During the compaction, we measure the density $\rho(t)$ of particles,
the density $\rho_{mob}(t)$ of mobile particles, the mobility
$\chi(t_w,t_w+t)=\frac{1}{dN}\sum_{a} \sum_{k=1}^{N} \frac{\delta
\left\langle (r_k^a(t_w+t) - r_k^a(t_w)) \right\rangle}{\delta f}$
obtained by the application of a random force to the particles between
$t_w$ and $t_w+t$, and the mean square displacement
$B(t_w,t_w+t)=\frac{1}{dN}\sum_{a} \sum_{k=1}^{N} \left\langle
(r_k^a(t_w+t)-r_k^a(t_w))^2 \right\rangle$ ($N$ is the number of
particles; $a=1,\cdots,d$ runs over the spatial dimensions: $d=2$ for
Tetris, $d=3$ for KA; $r_k^a$ is measured in units of the bond size
$d$ of the square lattice).  Indeed, the quantities $\chi(t_w,t_w+t)$
and $B(t_w,t_w+t)$, at equilibrium, are linearly related (and actually
depend only on $t$ since time-translation invariance holds) by
\begin{equation}
2 \chi(t) = \frac{X}{T^{eq}_d} B(t),
\label{FDT}
\end{equation}
where $X$ is the so-called Fluctuation-Dissipation ratio (FDR) which
is unitary in equilibrium. Any deviations from this linear law signals
a violation of the Fluctuation-Dissipation Theorem (FDT). Nevertheless
it has been shown, first in mean-field models~\cite{CuKu}, then in
various numerical simulations of finite dimensional
models~\cite{parisi,barratkob} how in several aging systems violations
from (\ref{FDT}) reduce to the occurrence of two regimes: a
quasi-equilibrium regime with $X=1$ (and time-translation invariance)
for ``short'' time separations ($t \ll t_w$), and the aging regime
with a constant $X \le 1$ for large time separations. This second
slope is typically referred to as a dynamical temperature $T_{dyn} \ge
T^{eq}_d$ such that $X = X_{dyn} = T^{eq}_d/T_{dyn}$~\cite{CuKuPe}.

We have simulated lattices of linear size $L=50,\,100,\,200$, in order
to ensure that finite-size effects were irrelevant. We have chosen
periodic boundary conditions on the lattice, having checked that other
types of boundary conditions (e.g. closed ones) gave the same
results. We have investigated the irreversible compaction dynamics of
the system up to times of $2 \times 10^5$ MC steps, realizing a large
number of different runs ($N_{runs} \simeq 8000 \div 9000$), in order
to obtain clean data. The random perturbation is realized by varying
the diffusion probability of each particle from the initial value
$p=\frac{1}{4}$ to the value $p^{\epsilon}= \frac{1}{4}+f_i^r \cdot
\epsilon$, where $f_i^r = \pm 1$ is a random variable associated to
each grain independently for each possible direction ($r=x,y$), and
$\epsilon$ represents the perturbation strength. The results presented
here are obtained with a perturbation strength $\epsilon=0.005$,
having checked that for $0.002< \epsilon < 0.01$ non-linear effects
are absent.

\subsection{Interrupted aging regime}

\begin{figure}[htb] 
\centerline{\psfig{file=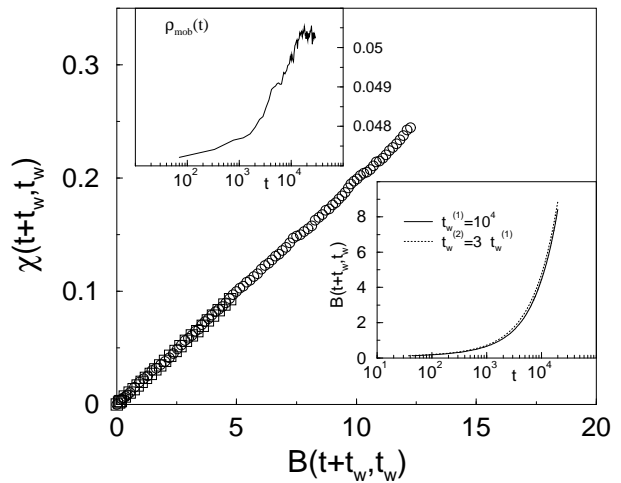,width=8.0cm}} 
\vspace{0.2cm} 
\caption{Einstein relation in the Tetris model with interrupted aging,
at various densities. Insets: Density of mobile particles vs. 
$t$, for  $t_w=10^4$, and $B(t_w+t,t_w)$ vs. $t$ for two different
$t_w$ (illustration of TTI).  }
\label{fig:FDT+TTI}
\end{figure}

When the compaction process is stopped at a certain time $t_w$, the
system relaxes toward equilibrium: the mean square displacement and
the integrated response function satisfy the time translational
invariance (TTI), as Fig.~\ref{fig:FDT+TTI} shows. This is the
so-called regime of interrupted aging, characterized by an increase of
$\rho_{mob}(t)$ toward its equilibrium value (inset of
Fig.~\ref{fig:FDT+TTI}) and a single linear relation for the $\chi$
vs. $B$ parametric plot. Fluctuation-Dissipation Theorem is then
recovered, and the value of the equilibrium fluctuation-dissipation
ratio, $T_d^{eq}$, is obtained from the measures of those quantities;
its value actually does not depend on the density of the system.

It is interesting to mention that the behavior described in this
subsection can actually also be observed for the KA model, which was
studied in~\cite{bakulose1,bakulose2} only under continuous
compaction.

\subsection{Aging dynamics}

\begin{figure}[htb] 
\centerline{
\psfig{file=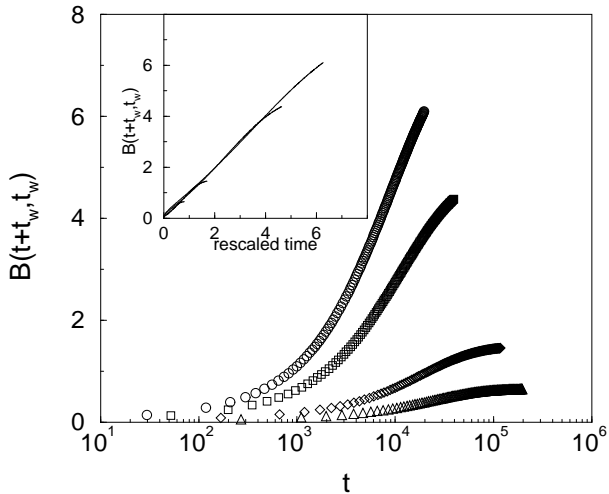,width=8cm}}
\vspace{0.2cm} 
\caption{Evolution of $B(t+t_w,t_w)$ for various $t_w$ (from bottom to
top, $5 \times 10^3$, $10^4$, $3 \times 10^4$ and $5 \times 10^4$);
inset: collapse of $B$; for the collapse, we have not taken into
account the final portion of each curve, because of a saturation of
$B$ due to finite-size effects.}
\label{fig:B}
\end{figure}

If the compaction is not stopped at time $t_w$, the density increases,
and a typical aging behaviour is observed, as shown in
Fig.~\ref{fig:B} where different curves of $B(t+t_w,t_w)$, for
different values of $t_w$, are reported: the mean square displacement
depends explicitly not only on the observation time $t$ but also on
$t_w$. The system remains out of equilibrium on all the observed time
scale, with the violation of time translational invariance (TTI). The
inset of Fig.~\ref{fig:B} shows the collapse of the curves of
$B(t+t_w,t_w)$ for different values of the age of the system, obtained
with the following scaling function
\begin{equation}
B(t+t_w,t_w)\,=\,c \cdot \left[ \frac{\ln \left(
\frac{t+t_w+t_s}{\tau} \right)}{ \ln \left( \frac{t_w+t_s}{\tau}
\right)} -1 \right]
\end{equation}
where $c,\,t_s$ and $\tau$ are fit parameters. Our results show a
linear dependence of the parameters $t_s$ and $\tau$ on the age of the
system, while the coefficient $c$ is nearly constant.

Moreover, we observe the density of mobile particles $\rho_{mob}(t)$
getting smaller than the corresponding value at
equilibrium~\cite{bakulose2}, as another evidence of the out of
equilibrium behaviour of the system during the compaction process.

\begin{figure}[htb] 
\centerline{\psfig{file=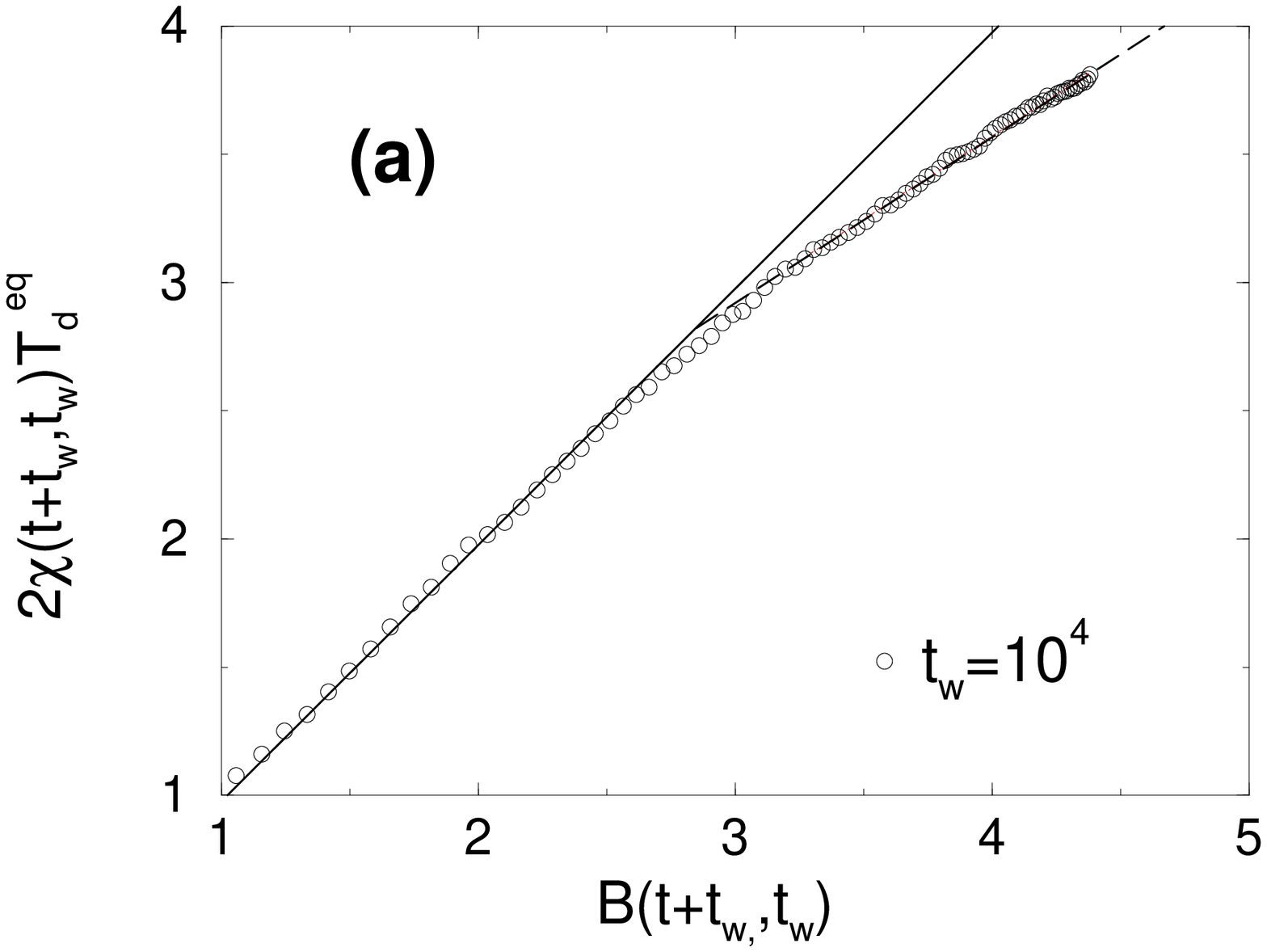,width=8.0cm}}
\centerline{\psfig{file=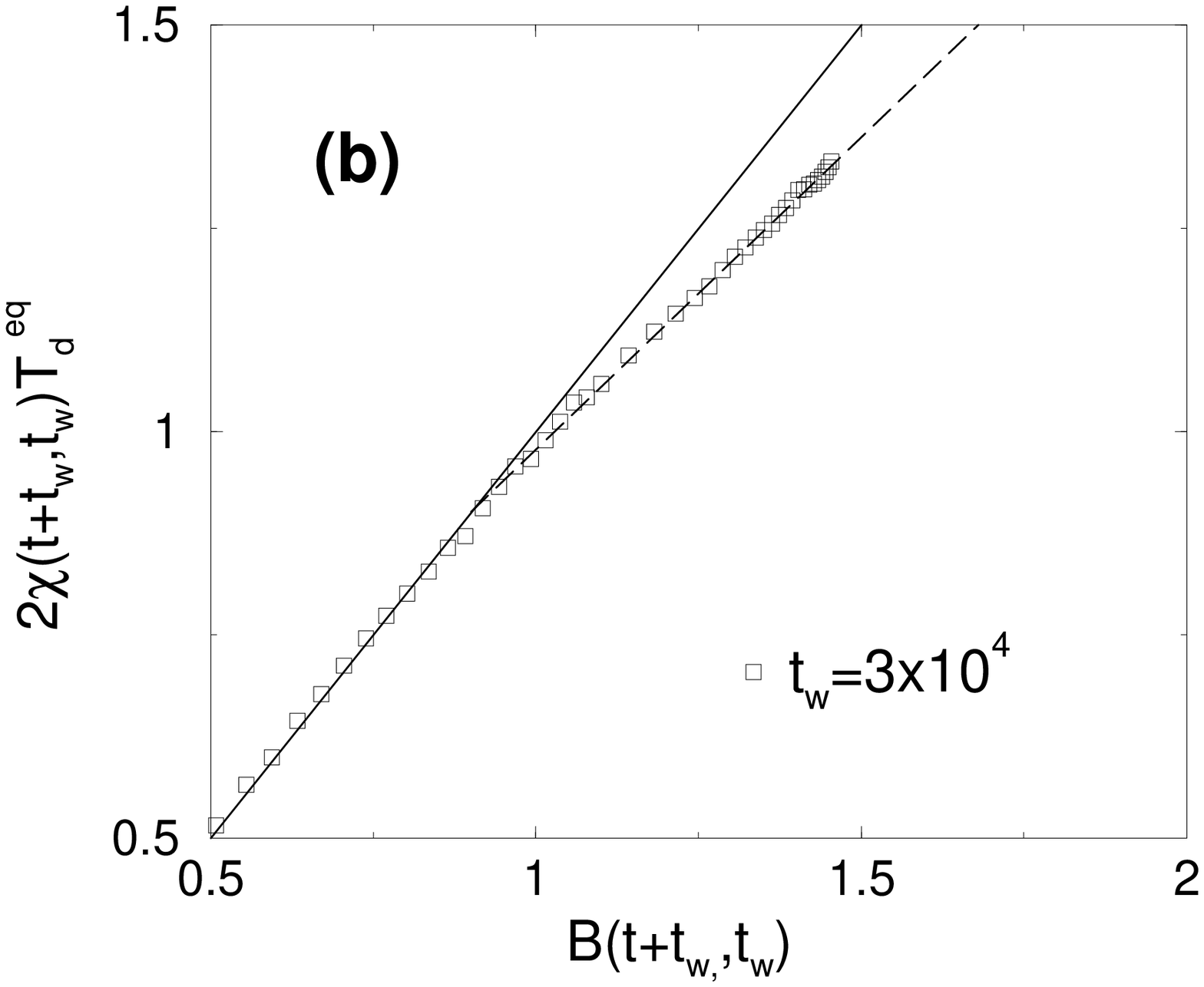,width=8.0cm}}
\vspace{0.2cm} 
\caption{Einstein relation in the Tetris model: plot of the mobility
$2 T_d^{eq}\chi(t_w,t_w+t)$ vs. the mean-square displacement
$B(t_w,t_w+t)$, for (a) $t_{w}^1=10^4$ and (b) $t_{w}^2=3 \times 10^4$. The
value of $T_d^{eq}$ is taken from Fig.~\ref{fig:FDT+TTI}, and the
slope of the full straight line is one. The dashed lines are linear of
the FDT violation whose slope gives a measure of $X_{dyn}$.}
\label{fig:x}
\end{figure}

The system also features a violation of the Fluctuation-Dissipation
Theorem. More precisely, the $\chi$ vs. $B$ parametric plot, reported
in Fig.~\ref{fig:x}, shows two different linear behaviours: for times
$t$ smaller than $t_w$ we observe a first quasi-equilibrium regime
where FDT holds (i.e. $X_{dyn}=1$), followed by a second regime in
which FDT breaks down and a dynamical temperature $T_{dyn}$ arises
which is independent of the observation time $t$. This quantity
actually depends on the age of the system, $t_w$, and therefore on the
density.  We have investigated this behaviour in a large range of
waiting times, corresponding to several values of the density,
obtaining the following results: $X_{dyn}^1=0.646 \pm 0.002, \,
X_{dyn}^2=0.767 \pm 0.005, \, X_{dyn}^3=0.784 \pm 0.005$ at
$t_w^1=10^4, \, t_w^2=3 \times 10^4$ and $t_w^3=5 \times 10^4$.

\subsection{Results for a bi-disperse system}

In order to investigate the dependence of the dynamical temperature on
the observables considered for its definition, we have introduced two
different types of particles.  Besides the ``T''-shaped particles,
already seen at the beginning of the section, we have considered
``L''-shaped particles. Such grains are characterized by a different
degree of disorder, so we expect these particles, the ``smaller''
ones, to move more easily, being less constrained. Although we have
two different types of particles, the system remains homogeneous
during compaction.

\begin{figure}[htb] 
\centerline{ \psfig{file=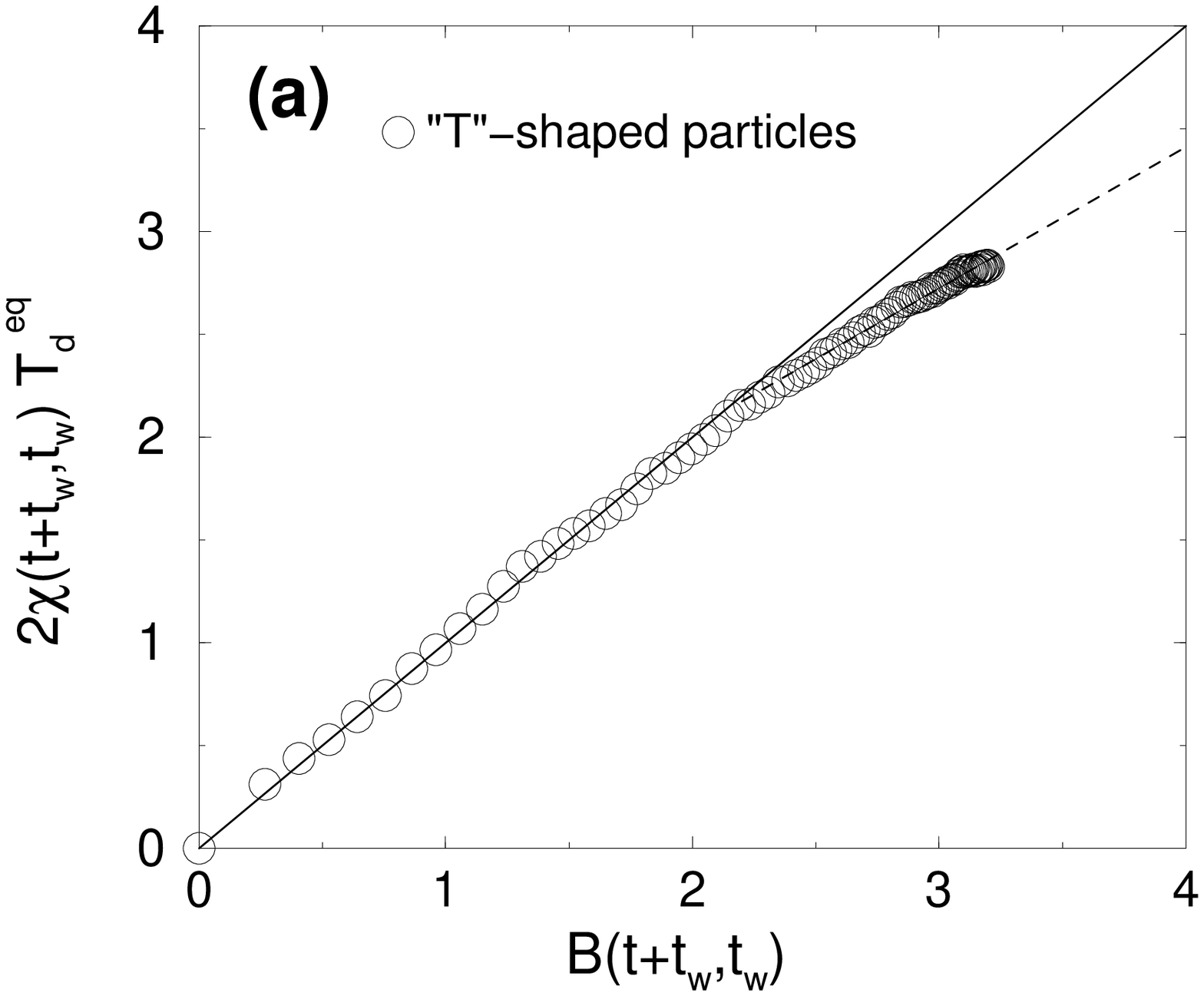,width=8.0cm}} \centerline{
\psfig{file=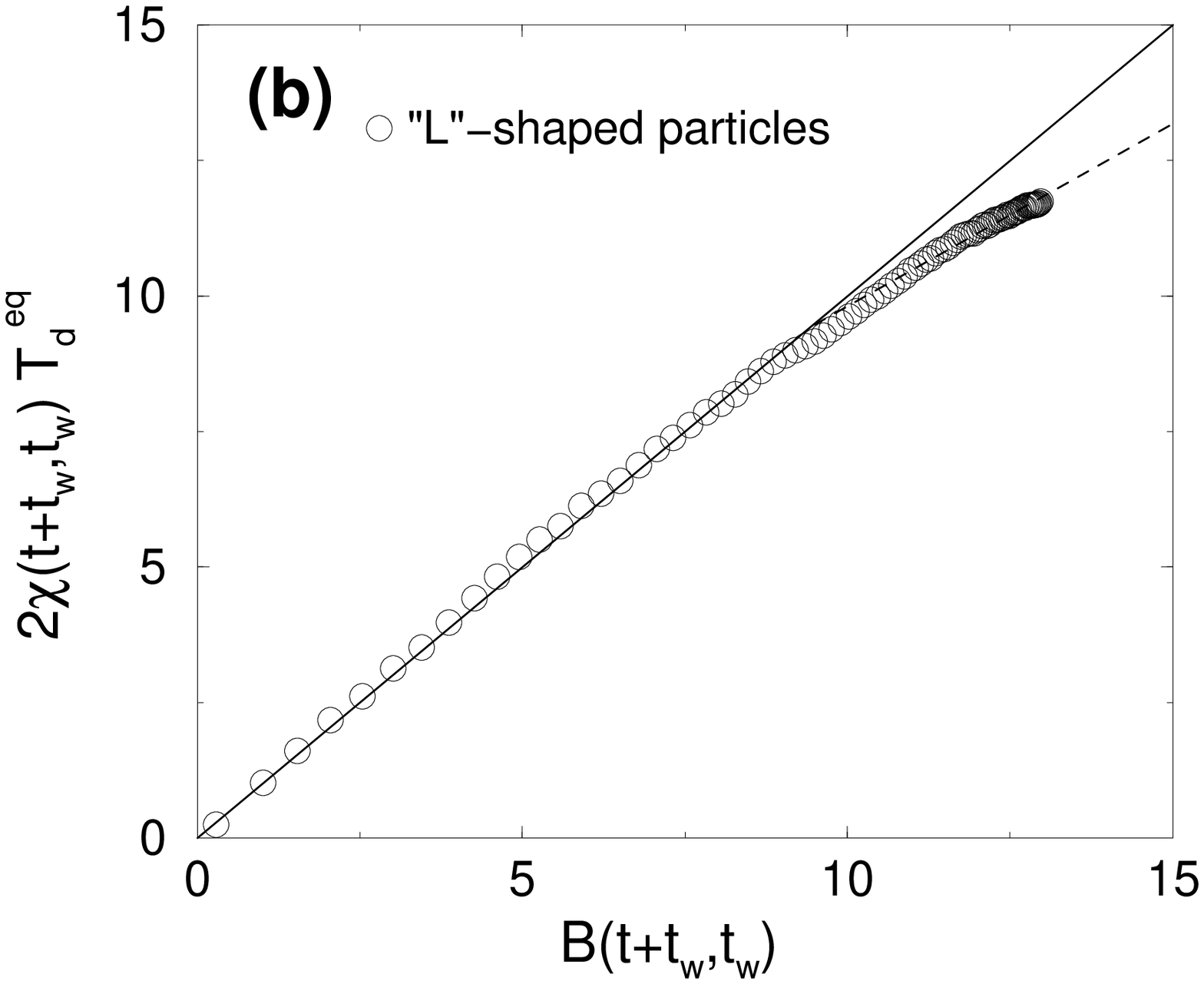,width=8.0cm}}
\vspace{0.2cm} 
\caption{Plot of the mobility $2 T_d^{eq} \chi(t_w,t_w+t)$ vs.  the
mean-square displacement $B(t_w,t_w+t)$, for different types of
particles (with equal concentrations) : (a) ``T''-shaped particles and
(b) ``L''-shaped particles, $t_w=5 \times 10^4$. The dynamical
temperatures associated to the two different types of particles
(i.e. the slopes of the dashed straight lines) are equal within the
error-bars.}
\label{fig:x2}
\end{figure}

We have therefore measured FDR for the two different types of
particles.  The results obtained show that the two dynamical
temperatures are equal within the error-bars, even though the related
diffusivities are different (see Fig.~\ref{fig:x2} for an example with
equal fractions of $T$-shaped and $L$-shaped particles). This result
has several important consequences. First of all the coincidence of
the results for the dynamical temperature obtained with different
observables is a crucial step for the establishment of a
thermodynamical interpretation. Another important consequence arises
from the experimental point of view: since the value of FDR is
independent from the shape of the particles, it could be possible to
measure the dynamical temperature of a granular material using a
tracer particle different from the particles composing the system.
This dynamical temperature should not depend on the shape of the
tracer particle. The extension of this result to the case of
compaction with gravity will be discussed in the next section.

\subsection{Comparison with Edwards' measure}

\begin{figure}[htb] 
\centerline{\psfig{file=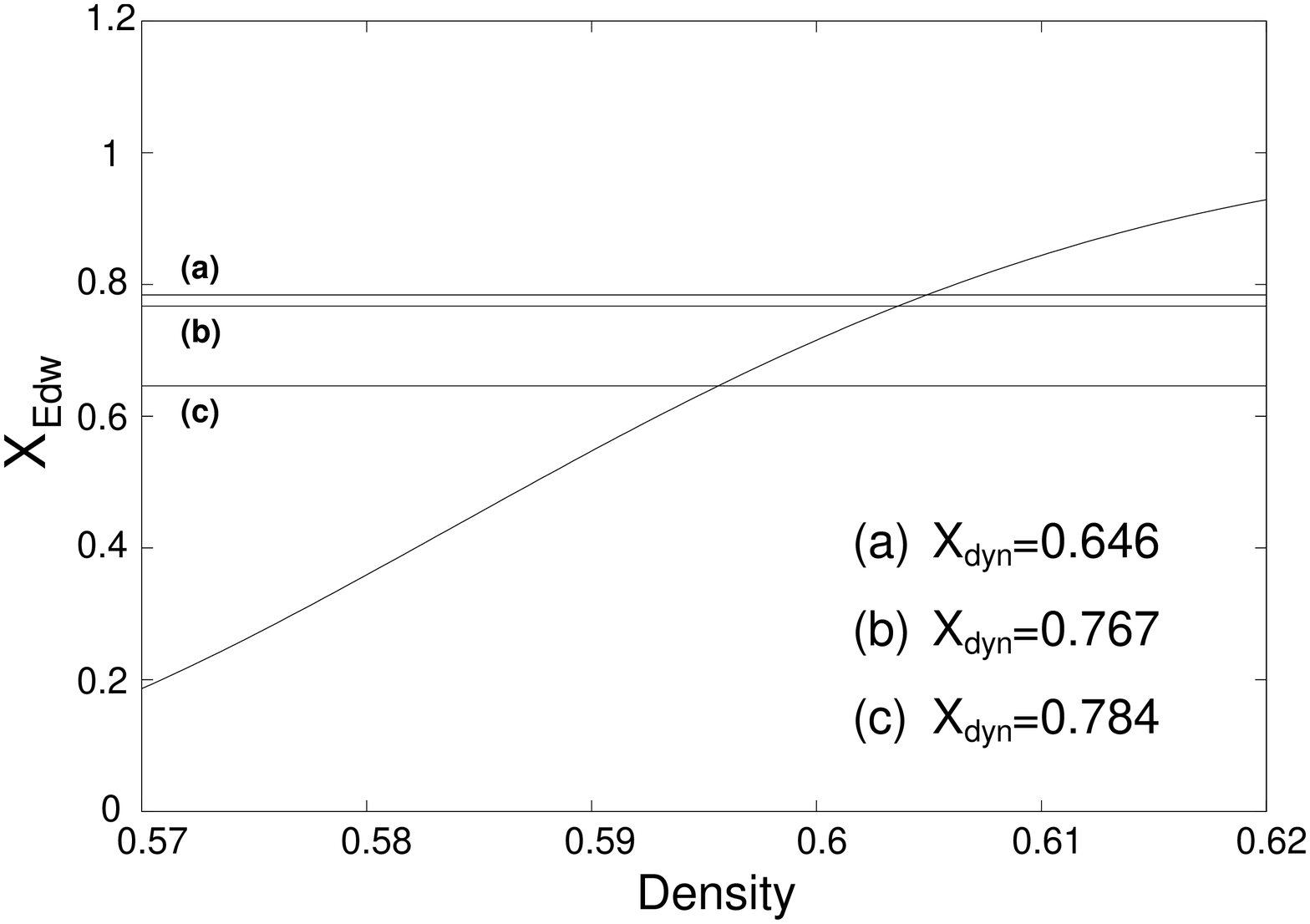,width=8.0cm}} 
\vspace{0.2cm} 
\caption{Static ratio $X_{Edw}$ as a function of density. The
horizontal lines correspond to the dynamical ratios $X_{dyn}$.
measured at $t_w=10^4,\ 3 \times 10^4,\ 10^5$ and determine the values
$\rho_1 \approx 0.596$, $\rho_2 \approx 0.603$, $\rho_3 \approx
0.605$, to be compared with Fig.~\ref{fig:rhox}.  }
\label{fig:comp} 
\end{figure}

\begin{figure}[htb] 
\centerline{\psfig{file=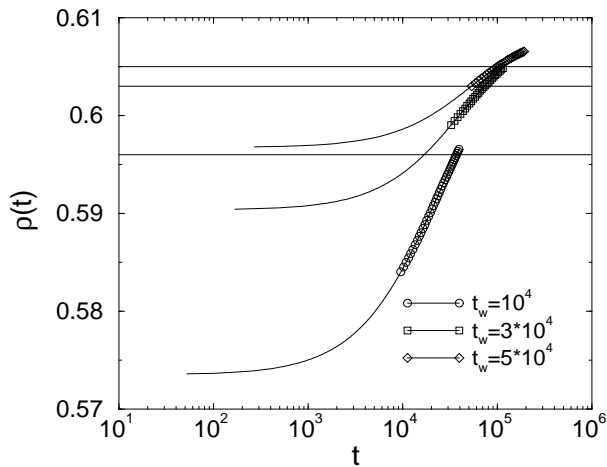,width=8.0cm}} 
\vspace{0.2cm} 
\caption{Evolution of the density during the measurements of 
$\chi$ and $B$, for $t_w=10^4,\ 3 \times 10^4,\ 10^5$. The evolution during
the quasi-equilibrium part is plotted with lines, and during the
violation of FDT with symbols. The horizontal lines correspond to
the densities $\rho_1$, $\rho_2$, $\rho_3$ from fig.~\ref{fig:comp}.
} 
\label{fig:rhox} 
\end{figure}

We are now in a position to compare the results for the
Fluctuation-Dissipation Ratio $X_{dyn}=T_d^{eq}/T_{dyn}$, measured
during the compaction dynamics, with the outcomes of Edwards' approach
at the corresponding densities.  In Fig.~\ref{fig:comp} we report the
values of $X_{Edw}$ vs. $\rho$ (as obtained with the Edwards' measure)
and $X_{dyn}$ (obtained by the FDR in dynamical measurements) at three
different values of $t_w$. In order to check the matching between
$X_{Edw}$ and $X_{dyn}$ it is enough to compare the densities obtained
from Fig.~\ref{fig:comp} by imposing $X_{Edw}= X_{dyn}$ with the
corresponding dynamical densities obtained at the corresponding
$t_w$. From Fig.~\ref{fig:comp} one gets: $\rho_1 \approx 0.596$ for
$t_{w}^1=10^4$, $\rho_2 \approx 0.603$ for $t_{w}^2=3 \times 10^4$,
$\rho_3 \approx 0.605$ for $t_{w}^3=5 \times 10^4$. On the other hand,
the evolution of the density of the system during the measurements of
the FDR is reported in Fig.~\ref{fig:rhox}. Since the measurements are
performed {\em during} the compaction, the density is evolving, going
from $\rho(t_w)$ to $\rho(t_w +t_{max})$. In each case, we obtain that
indeed $\rho_i \in [\rho(t_{w}^i), \rho(t_{w}^i +t_{max})]$, where we
have denoted with $\rho_i$ the densities obtained from
Fig.~\ref{fig:comp} for different values of $t_w^i$ ($i=1,2,3$).  More
precisely, $\rho_i$ is very close to $\rho(t_{w}^i +t_{max})$. This is
to be expected since the measure of the FDT violation is made for
times much larger than $t_{w}^i$ and, since the compaction is
logarithmic, the system actually spends more time at densities close
to $\rho(t_{w}^i +t_{max})$ than to $\rho(t_{w}^i)$.

\section{Fluctuation-Dissipation Ratio for the case with gravity}
\label{sec:V}

\subsection{Compaction dynamics}

While the use of a compaction without gravity is useful to study an
idealized context, real compaction due to shaking occurs because of
gravity.

The standard way of simulating the effect of gravity in a lattice
model is let the particles diffuse on a tilted (square or cubic)
lattice, with probabilities $p_{up}$ (resp. $p_{down}=1-p_{up}$) to go
up (resp. down), respecting the geometrical or kinetic constraints. A
closed boundary is situated at the bottom of the simulation box, of
horizontal linear size $L$ and vertical size $L_z >> L$ (lateral
boundary conditions can be closed or open, and various aspect ratios
can be used, without changing qualitatively the results). The control
parameter is the ratio $x=p_{up}/p_{down} < 1$.

This corresponds in fact to the dynamics at temperature $T=-1/\ln(x)$
for the Hamiltonian
\begin{equation}
H=\sum_i z_i \ ,
\label{eq:H}
\end{equation}
where the $z_i$ are the heights of the particles (of either of the
models under consideration) above the bottom.  Indeed, attempted
diffusion moves which respect the constraints are accepted with
probability $\min(1,x^{\Delta H})$.

As a result, particles tend to diffuse more easily toward the bottom.
A non-zero value of $x$ is however needed in order to allow for
rearrangements.

A simple lattice gas (with the only constraint of single occupancy)
diffusing with the above rule displays an equilibrium behaviour, with
the known density profile
\begin{equation}
\rho(z)= \frac{1}{1+\exp(\beta (z-z_0))} \ ,
\end{equation}
(with $\beta=1/T$ and $z_0$ depends on the number of particles) and
time translation invariance, and FDT is obeyed. The system is
therefore stationary and no evolution of the density occurs.

On the other hand, systems of constrained particles like the Tetris or
Kob-Andersen models are unable to reach this stationary state and are
stuck at lower densities (larger potential energies), with slow
compaction and aging, reproducing the phenomenology of granular
compaction~\cite{prltetris,BaLo,BaLo2,SeAr}. In particular, it has
been shown, both experimentally and numerically, that, due to
heterogeneities, the value of the potential energy (or of the bulk
density) is not the only relevant parameter~\cite{Josserand,BaLo2},
and that, in order to explain for instance memory phenomenon, it is
necessary to take into account the whole density profile along the
vertical direction. For example, various density profiles can be
obtained at approximately the same potential energy, by varying the
evolution of the forcing $T$ with time.

\begin{figure}[htb] 
\centerline{\psfig{file=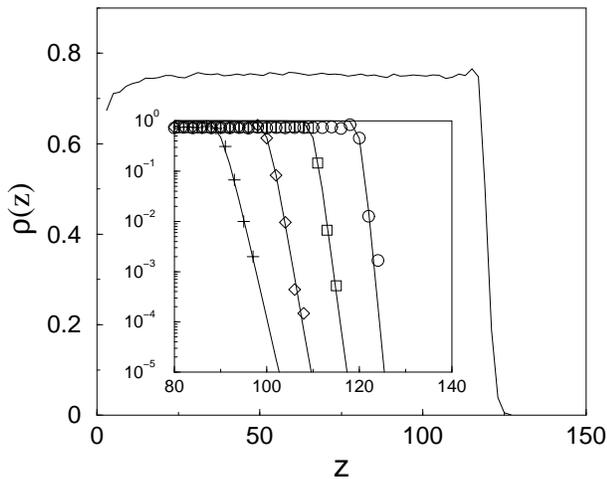,width=8.0cm}}
\vspace{0.2cm} 
\caption{Typical density profile obtained during the compaction
dynamics of the KA model.  The inset shows the dynamical interface for
$x=0.1$ (circles), $x=0.2$ (squares), $x=0.3$ (diamonds), and $x=0.4$
(crosses) together with equilibrium interfaces of a lattice gas
without constraints under gravity at the same shaking amplitudes
(lines).  The interface profiles have been horizontally shifted for
clarity.  }
\label{fig:profdyn}
\end{figure}

A first order treatment consists in separating a slowly compacting
bulk part and an interface~\cite{BaLo}. Because the interface is much
more dilute than the bulk, the particles feel much less the
constraints, and it turns out that the density profile at the
interface is exactly the same than the density profile of a lattice
gas without constraints (see the inset of Fig.~\ref{fig:profdyn}),
with Hamiltonian~(\ref{eq:H}) and forcing $T$. This part of the system
can therefore a priori be considered as ``in equilibrium'', i.e.  its
shape and dynamics are simply linked to the forcing.

At this level of treatment, the system is therefore considered to be
homogeneous in the horizontal directions, and heterogeneities are
taken into account only in the vertical direction.

Compaction data, under the effect of gravity, for various types of
Tetris model can be found in~\cite{prltetris,BaLo,BaLo2} and for the
KA model in~\cite{SeAr}.  Although we have monitored the usual
quantities describing the compaction, we will therefore not repeat
this analysis, but concentrate on the violation of FDT during
compaction, showing {\em en passant} that the few existing data can be
misleading or misinterpreted.

The existence of heterogeneities along the vertical direction moreover
leads to the following remarks:
\begin{itemize}
\item while calculating the Edwards' measure, imposing only the
potential energy of the system will lead to a unique density profile.
Since the dynamical density profile depends on the history, it is
already clear that specifying only the energy will not be sufficient
to predict all dynamical observables.

\item heterogeneities exist only along the vertical direction, so that
observables along the vertical and horizontal directions should a priori
be treated separately.

\item dynamical measures can either be made over the whole system or
restricted to the bulk. In the first case, the interface will
obviously give an ``equilibrium'' contribution that may be much larger
than the bulk contribution. 
\end{itemize}

Our numerical results have been performed in the following conditions.
\begin{itemize}

\item Tetris model: Horizontal size $L=40$ Number of particles
$N_{part}=1600$ number of runs $N_{runs}=2000$.

\item Kob-Andersen model: Horizontal size $L=20-40$, Number of particles
$N_{part}=30000-50000$, number of runs $N_{runs}=20-50$.

\end{itemize}

\subsection{FDT and its violations in the
existence of a preferred direction}

In general, the fluctuation-dissipation theorem relates, for a system
at equilibrium, conjugated response and correlation functions:
$$
T R(t,t') = \frac{\partial C(t,t')}{\partial t'} \ .
$$
Integrated between $t_w$ and $t+t_w$ in order to use the integrated
response function, the relation becomes
\begin{equation}
T\chi(t+t_w,t_w) = C(t+t_w,t+t_w) - C(t+t_w,t_w) \ .
\end{equation}

Let us first consider the case of horizontal degrees of freedom: in
these directions the system is homogeneous, and without drift.  Then
if we define
\begin{eqnarray}
&C_h&(t+t_w,t_w) = \frac{1}{2N}\sum_{a=x,y} \sum_{i=1}^{N} 
\left\langle a_i(t_w+t) a_i(t_w) \right\rangle \nonumber \\
&-&  \frac{1}{2} \sum_{a=x,y} \left\langle 
\frac{1}{N}\sum_{i=1}^{N}  a_i(t_w)\right\rangle
\left\langle 
\frac{1}{N}\sum_{i=1}^{N}  a_i(t_w+t) \right\rangle
\end{eqnarray}
and
\begin{equation}
B_h(t+t_w,t_w)=\frac{1}{2N}\sum_{a=x,y} \sum_{i=1}^{N} 
\left\langle (a_i(t_w+t)-a_i(t_w))^2 \right\rangle \ ,
\end{equation}
it is easily seen that
$B_h(t+t_w,t_w)=2 (C_h(t+t_w,t+t_w) - C_h(t+t_w,t_w))$.

Moreover, to measure susceptibilities, a perturbation is applied in
the following way: until $t_w$, the system evolves with forcing $x$
and Hamiltonian (\ref{eq:H}); at $t_w$, a copy is made and evolves
after $t_w$ according to the perturbed Hamiltonian $H_\epsilon= \sum_i
z_i^r + H_\epsilon^{h}$ where $H_\epsilon^h= \epsilon \sum_i (f_i
x_i^r + g_i y_i^r)$, with $f_i, g_i = \pm 1$ randomly for each
particle and $x_i^r,y_i^r,z_i^r$ are the positions of the particles in
the perturbed system. The integrated response
\begin{eqnarray}
\chi_h(t_w+t, t_w) = \frac{1}{2\epsilon N} \sum_{i=1}^N
\langle \, \overline{f_i  \left[ x_i^r(t_w+t) 
- x_i(t_w) \right]} \nonumber \\
+\overline{g_i\left[ y_i^r(t_w+t) - y_i(t_w) \right]} \,
\rangle 
\end{eqnarray}
can then be measured.

For a system at equilibrium (for example the simple lattice-gas with
single occupancy and no kinetic constraints, and Hamiltonian
(\ref{eq:H})), the FDT relation can be observed:
\begin{equation}
B_h(t_w+t,t_w)= 2 T \chi_h(t_w+t,t_w) \ .
\end{equation}
During compaction, the violation of FDT can then be investigated
from a parametric plot of $\chi_h$ vs. $B_h$. This is exactly similar
to the homogeneous case of section \ref{sec:IV}.

Up to now however, the only tentative measures of FDR have been
realized with observables coupled to the vertical
direction~\cite{nicodemi_response,Se2}.  This is in contrast with
other cases of systems with a preferential direction, where measures
along the only direction with no a priori heterogeneities were
undertaken~\cite{Makse,liu,ludo2}.

\begin{itemize}
\item
In~\cite{Se2}, the case of the KA model with a vertical random
perturbation was considered. The vertical mean square displacement
$$
B_v(t+t_w,t_w)=\frac{1}{N} \sum_{i=1}^{N} \left\langle
(z_i(t_w+t)-z_i(t_w))^2 \right\rangle
$$
was measured and confronted to the integrated response
$$
\chi_v(t_w+t, t_w) = \frac{1}{\epsilon N} \sum_{i=1}^N
\left\langle \, \overline{f_i  \left[ z_i^r(t_w+t) - z_i(t_w) \right]} \,
\right\rangle
$$
to a perturbation $H_\epsilon^v=\epsilon \sum_i f_i z_i^r$ ($f_i=\pm 1$
randomly). The existence of a dynamical
temperature was inferred from the observed linear relation between
$B_v$ and $\chi_v$, with a slope different from the applied
temperature.

\item
In~\cite{nicodemi_response}, a perturbation in the forcing was
applied, and confronted to the following mean-square displacement:
$$
\tilde{B}_v(t+t_w,t_w)=
\left\langle (h(t_w+t)-h(t_w))^2 \right\rangle 
$$
with $h(t)=\sum_i z_i(t)/N$. The perturbation in the forcing lead to
the observation of negative response functions
($\tilde{\chi}_v(t+t_w,t_w)= h^r(t+t_w)-h(t+t_w)$, where $h^r$ is the
mean height of the perturbed system, the perturbation being applied
after $t_w$), interpreted as the signature of a ``negative dynamical
temperature''. This case was investigated in~\cite{BaLo,BaLo2} where
this result was shown to be linked to the existence of memory effects,
as also confirmed in experiments~\cite{Josserand}.
\end{itemize}

In both cases however, the existence of a downward drift, due to
compaction, was not taken into account for the correct definition of
the correlation part of the fluctuation-dissipation relation: indeed,
in the first case, the correlation being
\begin{eqnarray}
&C_v&(t+t_w,t_w) = \frac{1}{N} \sum_{i=1}^{N} 
\left\langle z_i(t_w+t) z_i(t_w) \right\rangle \nonumber \\
&-&   \left\langle 
\frac{1}{N}\sum_{i=1}^{N}  z_i(t_w)\right\rangle
\left\langle 
\frac{1}{N}\sum_{i=1}^{N}  z_i(t_w+t) \right\rangle \ ,
\end{eqnarray}
$B_v(t+t_w,t_w)$ is not proportional to
$C_v(t+t_w,t+t_w)-C_v(t+t_w,t_w)$ as in the homogeneous case.
This is even more easily seen in the second case, where the correlation 
conjugated to the response to a change in the driving is
$\tilde{C}_v(t+t_w,t_w) = \langle h(t+t_w)h(t_w) \rangle - 
\langle h(t+t_w)\rangle \langle h(t_w) \rangle$.
Indeed 
$$
\tilde{B}_v(t+t_w,t_w)= \langle h^2(t+t_w) +h^2(t_w) \rangle
-2 \langle h(t+t_w)h(t_w) \rangle \ ,
$$
and 
\begin{eqnarray}
&\tilde{C}_v&(t+t_w,t+t_w) - \tilde{C}_v(t+t_w,t_w) = \nonumber \\ &
&\langle h^2(t+t_w) \rangle -\langle h(t+t_w)h(t_w) \rangle \nonumber
\\ &-&\langle h(t+t_w)\rangle \langle h(t+t_w) - h(t_w) \rangle
\label{eq:corcoherente}
\end{eqnarray}
are not simply related since
$\langle  h(t+t_w)\rangle \ne \langle  h(t_w)\rangle $
and $\langle h^2(t+t_w)\rangle \ne \langle h^2(t_w) \rangle$
(see also a similar discussion, on the case of one-time
quantities changing with time, in \cite{buhot}).

It turns therefore out that the results of
\cite{Se2,nicodemi_response} are a priori flawed from an incorrect
measure of the correlation part of FDR.

We will see in the next subsections how measures of correlation and
response functions along the horizontal directions lead to sensible
results, whereas all measures of vertical correlations or response
lead to the impossibility of defining effective temperatures.

\subsection{FDR in the aging (compacting) regime}

\subsubsection{Vertical observables?}

Two sets of response and correlation functions can a priori be
measured: the incoherent ones ($C_v, \chi_v$) as in \cite{Se2} or the
coherent ones ($\tilde{C}_v, \tilde{\chi}_v$) as in
\cite{nicodemi_response}.

If we write
\begin{eqnarray}
&C_v&(t+t_w,t+t_w)- C_v(t+t_w,t_w) = \nonumber \\ &=& \left\langle
\frac{1}{N} \sum_{i=1}^N z_i(t+t_w) (z_i(t+t_w)-z_i(t_w))
\right\rangle \nonumber \\ &+&\frac{1}{N} \left\langle \sum_{i=1}^N
z_i(t+t_w) \right\rangle \left\langle \frac{1}{N} \sum_{i=1}^N (
z_i(t_w) -z_i(t+t_w) ) \right\rangle \nonumber \\ &=&\left\langle
\frac{1}{N}\sum_{i=1}^N ( z_i(t_w) -z_i(t+t_w) ) \left( \left\langle
h(t+t_w) \right\rangle -z_i(t+t_w) \right) \right\rangle \nonumber
\end{eqnarray}
we can observe that generically $z_i(t_w+t) \le z_i(t_w)$ since the
system is compacting, so that two opposite contributions can be
distinguished in $C_v(t+t_w,t+t_w)- C_v(t+t_w,t_w)$: the particles
such that $z_i(t+t_w) < h(t+t_w)$ give a positive contribution, those
such that $z_i(t+t_w) > h(t+t_w)$ give a negative one. At short and
intermediate times, the particles closer to the surface move more than
those in the bulk and therefore the negative contribution dominates.
This leads to a negative $C_v(t+t_w,t+t_w)- C_v(t+t_w,t_w)$. At very
long times $C_v(t+t_w,t+t_w)- C_v(t+t_w,t_w)$ has to become positive
by definition, but such times may not be reachable in a numerical
simulation.

This peculiar behaviour comes from the fact that the drift is not
homogeneous in the system: some regions are compacting more than
others. Local drifts should then be taken into account. However this
is numerically (and also experimentally) too difficult to measure.

\begin{figure}[htb] 
\centerline{
\psfig{file=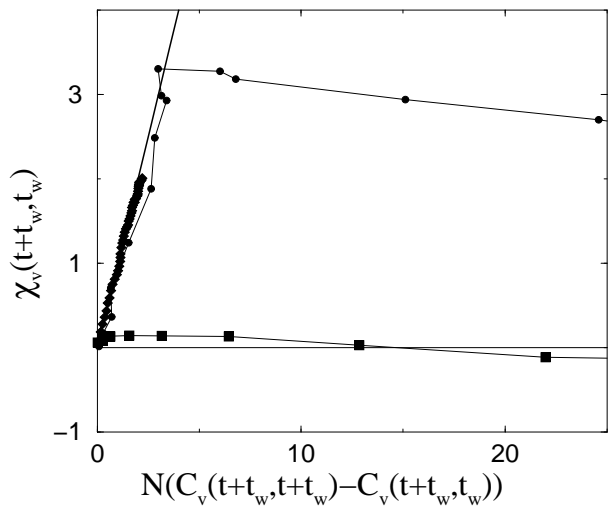,width=8.0cm}}
\vspace{0.2cm} 
\caption{$\tilde{\chi}_v (t+t_w,t_w)$ vs.  $N(\tilde{C}_v(t+t_w,t+t_w)
- \tilde{C}_v(t+t_w,t_w))$ for the KA model, with $L=20$,
$N_{part}=4000$, $N_{run}=500$, $x=0.8$ (circles) and $x=0.5$
(squares); $t_w=2^{14}$ and $t=2,\cdots,2^{16}$.
The first equilibrium part corresponds to the interface
dynamics, while at longer times $\tilde{\chi}_v$ decreases because of
the bulk response.  The diamonds correspond to a simulation with no
kinetic constraint and $N_{part}=4000$ particles: only equilibrium FDT is
then observed. The straight line has slope $1$.  }
\label{fig:pert_vert}
\end{figure}

On the other hand, the coherent correlation and response
$\tilde{C}_v$, $\tilde{\chi}_v$ can also be measured. The difficulty
arises from the measure of the coherent correlation function
$\tilde{C}_v$, of order $1/N$ for $N$ particles: relatively small
systems have to be simulated with a large number of realizations. For
an equilibrium lattice gas without kinetic constraints, FDT is then
recovered: $N(\tilde{C}_v(t+t_w,t+t_w) - \tilde{C}_v(t+t_w,t_w))=
\tilde{\chi}_v (t+t_w,t_w)$.  In the case of the compacting system,
the parametric plot of $N(\tilde{C}_v(t+t_w,t+t_w) -
\tilde{C}_v(t+t_w,t_w))$ versus $\tilde{\chi}_v (t+t_w,t_w)$ reveals a
first part of slope one, which corresponds to the fast, equilibrium
response of the interface.  At larger times $t$ however, the response
of the bulk, which can compactify more easily if the forcing is
increased, leads to a decrease of $\tilde{\chi}_v (t+t_w,t_w)$, which
can even become negative as observed in~\cite{nicodemi_response}.  As
$t_w$ goes to $\infty$, the bulk becomes so compact that its
contribution goes to zero, and the equilibrium FDT can be recovered
thanks to the interface contribution. Those results are summarized in
Fig.~\ref{fig:pert_vert}.

The previous investigations shows that no definition of an effective
temperature can be inferred from dynamical measures correlated with
the preferred directions in which heterogeneities occur.

Note that this kind of situation also arises in the study of effective
temperatures in driven vortex lattices with random pinning: while an
effective temperature can be defined and measured for degrees of
freedom perpendicular to the drive, problems are encountered when
dealing with longitudinal observables \cite{leto}.

We now turn to horizontal observables.

\subsubsection{Horizontal observables}

The first result is obtained by studying the whole system with a
horizontal perturbation applied: the relation between mean-square
displacement $B_h$ and response function $\chi_h$ is then clearly
linear, with a slope equal to the temperature $T$ of the forcing. This
seemingly surprising result is easily explained by the fact that both
functions are completely dominated by the contribution of the
interface where the particles can diffuse quite easily, and which
actually displays an equilibrium profile (see Fig.~\ref{fig:profdyn}).
It seems therefore natural to restrict the study of the observables to
the bulk part of the sample, in which the density is quite homogeneous
when a constant forcing is applied (see~\cite{SeAr,BaLo2} and
Fig.~\ref{fig:profdyn}).  The sums defining $B_h$, $\chi_h$ are
therefore restricted to the particles that remain between $t_w$ and
$t_w+t$ in the bulk (defined e.g. as $z_{min} < z_i < z_{max}$, with
$z_{min}$ and $z_{max}$ appropriately chosen).

\begin{figure}[htb] 
\centerline{\psfig{file=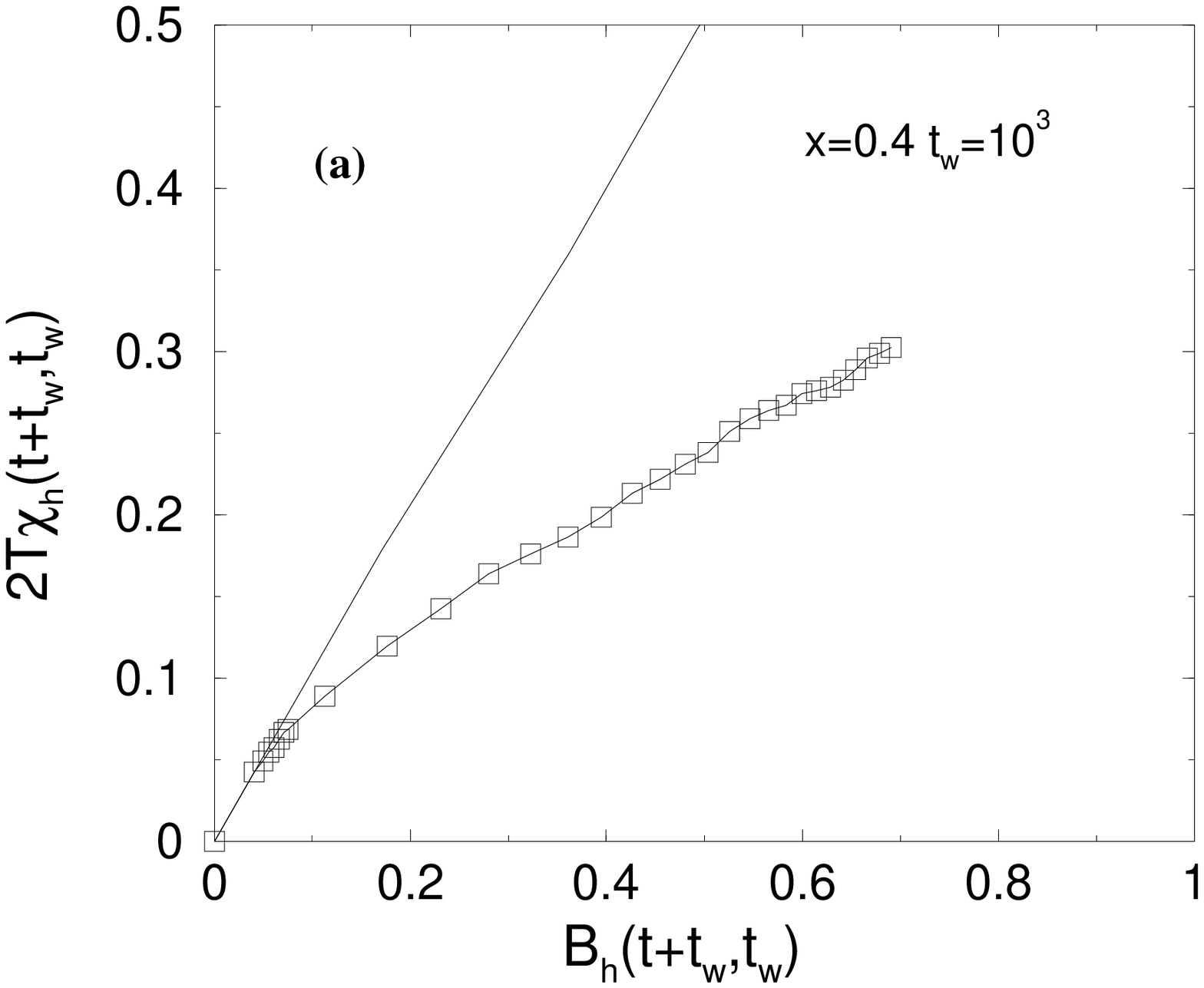,width=8.0cm}} \centerline{
\psfig{file=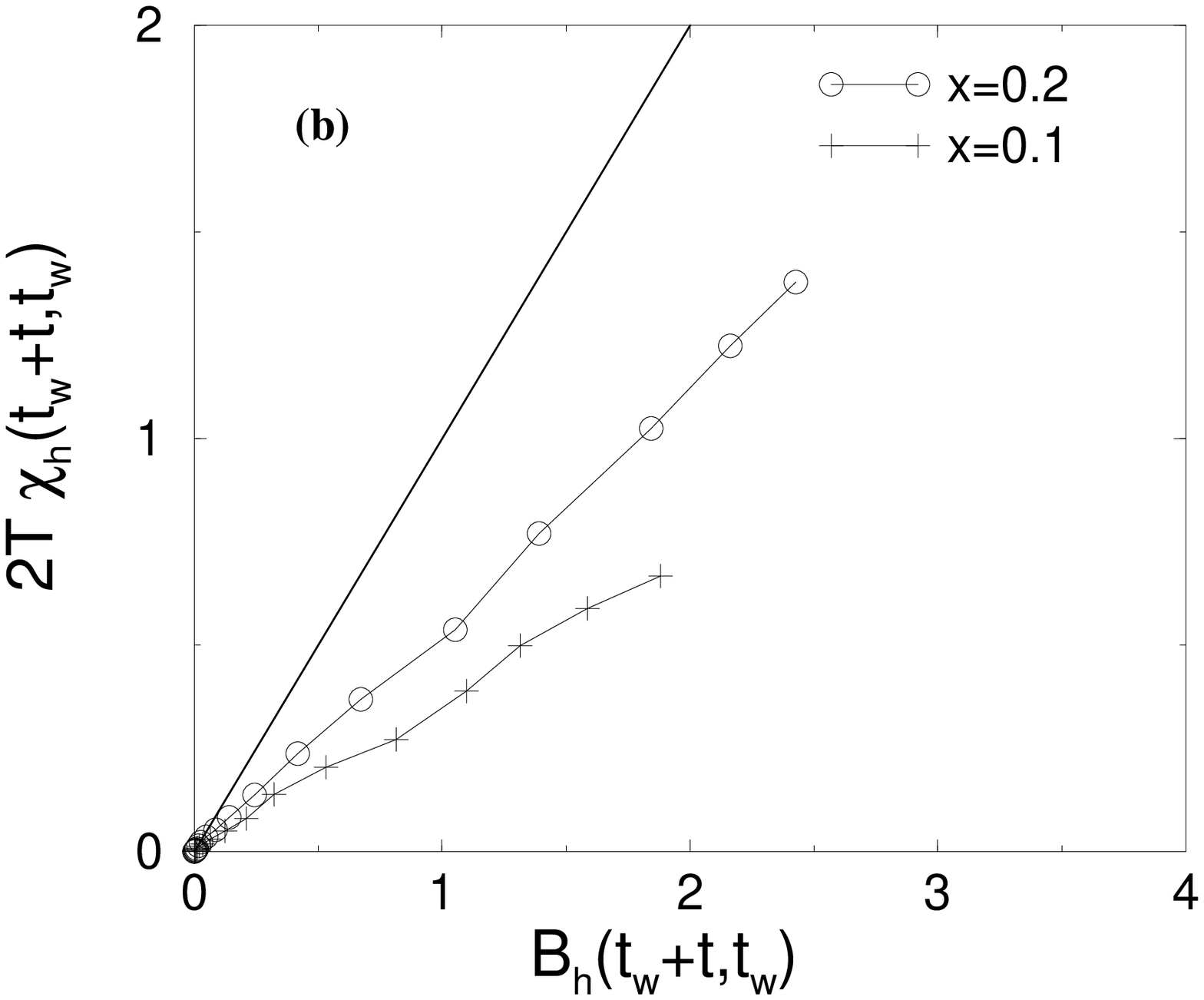,width=8.0cm}}
\vspace{0.2cm} 
\caption{$2 T \chi_h$ vs. $B_h$ for the Tetris model (a) and the
KA model ($t_w = 2^{14}$)(b). The straight lines have slope $1$.}
\label{fig:hor_vert_pert_tetris_ka}
\end{figure} 

Our results, summarized in Fig.~\ref{fig:pert_vert}
and~\ref{fig:hor_vert_pert_tetris_ka},
are qualitatively similar for both models. The clear violation of
FDT obtained with horizontal perturbations allows for the measurements
of the FD ratios while nothing can be said using the data obtained with
vertical perturbations.

\subsection{Results for a bi-disperse system}

In the models we have considered, it is quite easy to implement the
presence of two types of particles (this has already been seen in
section \ref{sec:IV} for the Tetris model without gravity).

For the Kob-Andersen model, we can simulate ``small'' and ``large''
particles by taking different values for the kinetic constraint, e.g.
$\nu_1=5$ and $\nu_2=6$ or $\nu_2=7$. As also shown in~\cite{SeAr},
partial segregation then occurs because the particle with larger $\nu$
are less constrained and can move more easily toward the bottom.

\begin{figure}[htb]
\centerline{\psfig{file=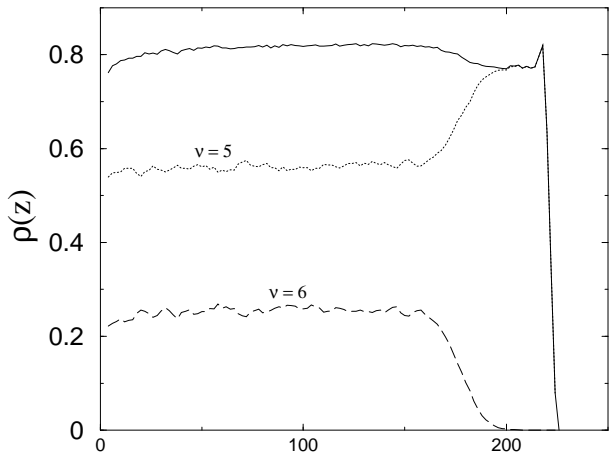,width=8.0cm}}
\vspace{0.2cm}
\caption{Density profiles for a bidisperse system of $30000$
particles with $\nu_1=5$ and $10000$ particles with $\nu_2=6$, for
a forcing $x=0.2$, after $2^{14}$ time steps. The less constrained 
particles have diffused more easily towards the bottom.}
\label{fig:prof2}
\end{figure}

However, as shown in Fig.~\ref{fig:prof2}, there exists a bulk region
in which the density profiles for both types of particles are flat. It
is therefore possible to measure FDR in this region.

\begin{figure}[htb]
\centerline{\psfig{file=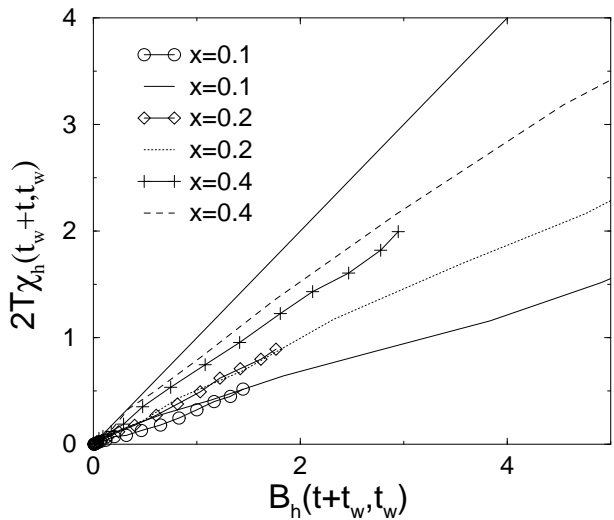,width=8.0cm}}
\vspace{0.2cm}
\caption{KA model: $2 T \chi_h$ vs. $B_h$, measured in the homogeneous
bulk, for the two types of particles (symbols: more constrained
particles with $\nu_1=5$; lines: less constrained particles with
$\nu_2=6$), for various forcing ($0.1$, $0.2$ and $0.4$) and
$t_w=2^{14}$, $t=2,\cdots,2^{18}$. The two kinds of particles display
different $\chi_h$ and $B_h$ at a given time, but the same violation
of FDT.  }
\label{fig:fdr2}
\end{figure}

The results, shown in Fig.~\ref{fig:fdr2}, are quite clear: although
the smaller particles are more mobile than the larger ones, and
therefore diffuse more easily, the FDR for the two types of particles
are equal. As already noted in section \ref{sec:IV}, this result is
important since it means that the FDR can in principle be measured
using tracers different from the particles composing the granular
material, and that the FDR should then be independent from the shape
of the tracer.

\subsection{Edwards' measure}

Edwards' measure is defined as a flat measure over all blocked
configurations, i.e. configurations with all particles unable to move.
For a system under gravity, a particle at height $z$ is `blocked' if
it cannot move downwards, i.e. if all its neighbouring sites at $z-1$
are occupied or if the particle cannot move toward either of these
sites because of the geometrical or kinetic constraints.

Implementing this difference (with respect to the case without
gravity) into the auxiliary model of~\cite{bakulose1,bakulose2} is
straightforward.

\begin{figure}[htb]
\centerline{\psfig{file=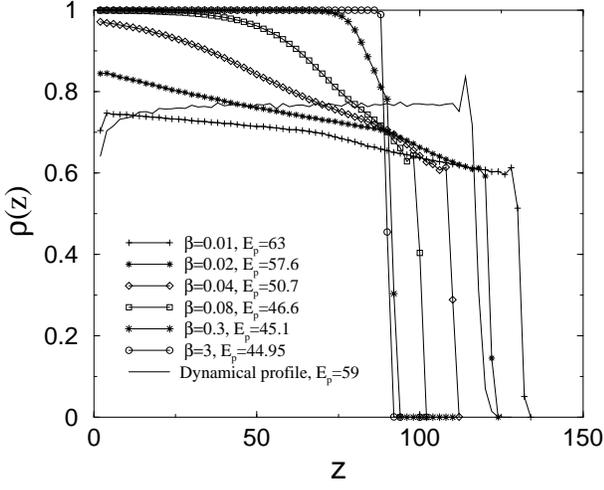,width=8.0cm}}
\vspace{0.2cm}
\caption{KA model: Density profiles for Edwards' measure, obtained with various
values of the Lagrange parameter $\beta$ (symbols).  A dynamically
obtained profile is also shown for comparison: it corresponds to a
constant shaking $x=0.4$ and to a potential energy similar to the case
$\beta=0.02$ (stars). The profiles are very different.  }
\label{fig:edw_prof}
\end{figure}

As in~\cite{Makse}, the following procedure is used: the auxiliary
model has total `energy' $\beta_{aux} E_{aux} + \beta E_p$ where
$E_{aux}$ is as usual the number of mobile particles, $\beta_{aux}$
the auxiliary inverse temperature, $E_p$ the potential energy and
$\beta$ a Lagrange multiplier.  For each value of $\beta$, an
annealing procedure is performed on $\beta_{aux}$, until
configurations with $E_{aux}=0$ are reached.  The density profiles are
then measured, along with the value of $E_p$.  Repeating the procedure
then yields the curve $\beta_{Edw} (E_p)$ directly.  The profiles at
various values of $E_p$ are shown in Fig.~\ref{fig:edw_prof}.  These
profiles are quite different from the dynamically obtained profiles at
similar energies. This is not surprising since they have been obtained
imposing only the potential energy, whereas the dynamical profiles
depend on the history and it has been
shown~\cite{Josserand,BaLo,BaLo2} that $E_p$ is not the only relevant
parameter.

In this case, Edwards' measure, if constructed by imposing only one
parameter, is not able to predict dynamical observables.

\begin{figure}[htb]
\centerline{\psfig{file=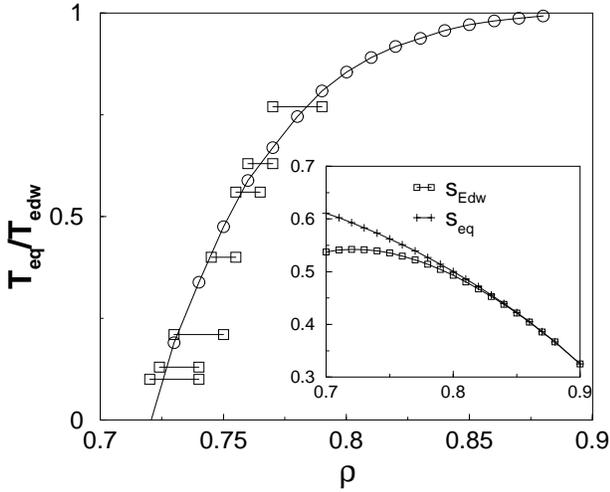,width=8.0cm}}
\vspace{0.2cm}
\caption{Circles: $X_{Edw} = \frac{ds_{equil}(\rho)}{d\rho} /
\frac{ds_{Edw}(\rho)}{d\rho} $ vs. $\rho$ for KA model, imposing
homogeneity; squares: dynamically obtained FDR (for horizontal
displacements and response) vs. density measured for the corresponding
dynamical profiles. Inset: equilibrium and Edwards' entropy densities
(imposing homogeneity for Edwards' entropy).  }
\label{fig:edw_hom}
\end{figure}

On the other hand, since dynamically the bulk density profiles are
flat, we can generate blocked configurations with homogeneous density,
at various densities. This yields a restricted Edwards' measure;
proceeding as in~\cite{bakulose1,bakulose2} we obtain Edwards' entropy
at the densities considered and we can therefore compute
\begin{equation}
X_{Edw} = \frac{ \frac{ds_{equil}(\rho)}{d\rho}}{
        \frac{ds_{Edw}(\rho)}{d\rho}}
\end{equation}
which is shown in Fig.~\ref{fig:edw_hom} in the case of the KA
model.

\subsection{Comparison and discussion}

From the dynamics on the one hand and Edwards' measure on the other
hand two sets of data are obtained:
\begin{itemize}
\item dynamical FDR at various densities, for horizontal displacements
and response functions; since the density is evolving during the
measures, an uncertainty is observed;
\item statically obtained $X_{Edw}$.
\end{itemize}

Fig.~\ref{fig:edw_hom} shows that the agreement between both sets of
data is very good, even for a quite large vibration $x=0.4$ or low
densities $0.73$.

\begin{figure}[htb]
\centerline{\psfig{file=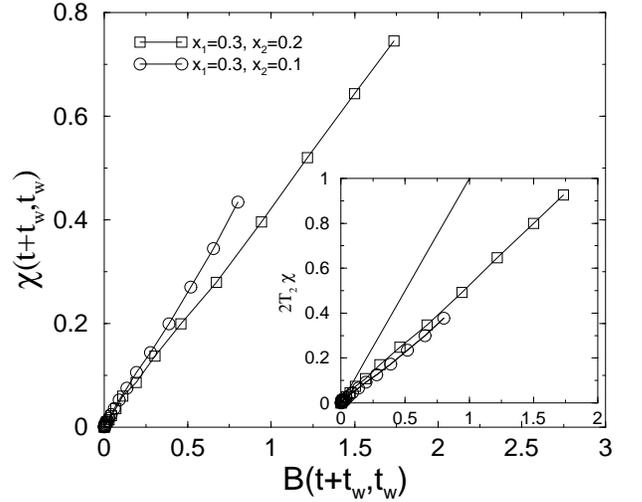,width=8.0cm}}
\vspace{0.2cm}
\caption{$\chi$ vs. $B$ following a change in the forcing from $x_1$
to $x_2$ at $t_w=2^{14}$: the slope depends on the forcing applied
after $t_w$ ($t=2,\cdots,2^{18}$). Inset: $T_2 \chi$ vs. $B$, showing
the equality of the dynamical ratios $T_2/T_{dyn}$.  }
\label{fig:x1x2}
\end{figure}

The theoretical results can be summarized as follows: in the case of a
homogeneous bulk, the ratio $X_d$ of the horizontal dynamical
temperature to the imposed temperature $-1/\ln(x)$ only depends on the
bulk density, and is given by $X_{Edw}(\rho)$.  Using various
vibration amplitudes, we have checked that $X_d$ at various densities
and $X_{Edw}(\rho)$ indeed coincide.  Another check of the consistency
of the theoretical construction can be made by comparing two dynamical
procedures: if a certain forcing $T_1$ is applied until $t_w$, and
then changed to $T_2$, a dynamical temperature $T_d^{12}$ will be
obtained.  While $T_d^{12}$ depends on $T_2$, the equality
$$
\frac{T_2}{T_d^{12}} = X_{Edw}(\rho(t_w)) 
$$
should be observed, at least if $T_2$ is not much higher than $T_1$
(in this case, as shown in~\cite{BaLo2}, the bulk density changes
suddenly by a large amount after $t_w$).  We have performed such
measurements and checked that this is indeed the case (see
Fig.~\ref{fig:x1x2}).

\section{Relations with experiments}
\label{sec:VI}

While the link between a dynamically measured temperature and a static
measure is of great theoretical importance, it is experimentally
impossible to sample such a measure. Theoretical predictions can be
checked experimentally only through purely dynamical measures.

Preliminary results in this direction were obtained by the Chicago
group~\cite{exp-chicago} by measuring the volume fluctuations with
respect to the steady-state volume at different heights of the
sample. Since the curves obtained at different heights do not coincide
the authors concluded that one single observable, namely the Edwards
compactivity, cannot account for the depth dependence of the
fluctuations. These conclusions are in agreement with our results
obtained in systems with a preferential direction where fixing one
single observable in the Edwards approach does not allow for the
prediction of the dynamical observables unless one reduces the
analysis to some homogeneous section of the system.

Thanks to recent theoretical progresses, new experiments have been
proposed in order to check the existence of dynamical
temperatures~\cite{bakulose1,Makse,ludo2,liu}, by monitoring mean
square displacements and mobility of tagged particles, or tracers, in
sheared super-cooled liquids or foams, or in sheared or tapped
granular media. The existence of a linear relation ($B$ vs. $\chi$)
could be tested for various shapes, masses, etc. of the tracers, in
order to check that this relation is indeed defining a temperature.

Our analysis of models compacting under gravity suggests also other
types of experimental possibilities. First, only diffusivity and
mobility in the direction perpendicular to the gravity should be
measured. Moreover, the existence of strong heterogeneities implies
that a tracer close to the interface should allow to measure a
temperature $T_d^i$ which depends directly on the driving amplitude
(in the models, $T=-1/\ln(x)$), and is stationary, i.e. does not
depend on the bulk density. On the other hand, a tracer well immersed
in the bulk should yield another value $T_d^b$ of the FDR (which
depends on $t_w$), and the {\em ratio} $T_d^b/T_d^i$ should depend
only on the density of the bulk. For example, if experiments are
performed changing the driving intensity at $t_w$ from $x_1$ to $x_2$,
and the two associated ``interface'' and ``bulk'' temperatures are
measured, the ratio $T_d^b/T_d^i$ should be independent of $x_2$
provided the bulk density does not change significantly. In this way,
the comparison of only dynamical measures would be a strong
experimental test for the whole theoretical construction, without any
need to sample the underlying static measure.

\section{Summary and Conclusions}
\label{sec:VII}

In this paper, we have studied two paradigmatic models for the
compaction of granular media. These models consider particles
diffusing on a lattice, with either geometrical or dynamical
constraints. Idealized compaction without gravity has been implemented
for the Tetris model, and compaction with a preferential direction
imposed by gravity has been studied for both models. The possibility
to define dynamically a temperature in the framework of
Fluctuation-Dissipation relations and to link it to the statically
constructed Edwards' measure has been investigated.

In the first, ideal, case of the homogeneous compaction, the obtained
FD ratio have been clearly shown to be in agreement with the
prediction of Edwards' measure at various densities.

The situation is more complicated if a preferential direction is
present: then the whole density profile has a priori to be taken into
account. Moreover, the vertical drift due to compaction leads to
contradictory (and sometimes meaningless) results when observables
coupled to the preferential direction are considered for the
evaluation of a FD ratio~\cite{nicodemi_response,Se2}. Since the
energy of the system is not the only parameter, and since the density
profiles depend on the history, Edwards' measure is not a priori able
to predict the dynamical configurations.  If however the homogeneity
of the bulk is imposed, FD ratio obtained dynamically for horizontal
displacements and mobility and from Edwards' measure coincide.

It is striking to note that Edwards' measure, which a priori could be
valid only for very weak forcing and almost stationary systems seems
however to yield good predictions even for non-stationary systems that
are still compacting.

Finally, we have proposed experimental tests of the whole theoretical
construction, through the comparison of various types of dynamical
measurements, since the construction of Edwards' measure, numerically
straightforward, is obviously impossible in experiments.  {\large
Acknowledgments} The authors wish to warmly thank an anonymous referee for
having pointed out an important inconsistency in a first version
of this work. This work has been
partially supported by the European Network-Fractals under contract
No. FMRXCT980183.

\end{document}